\documentclass[11pt,twoside]{article}

\usepackage{asp2006}
\usepackage{epsf}
\usepackage{lscape}
\usepackage{graphicx}

\begin{document}

\def\HI{H{\sc i} }
\def\HII{H{\sc ii} }
\def\Ha{H$\alpha$ }
\def\OIII{O[{\sc iii}] }
\def\Msun{M$_{\odot}$ }
\def\eps{$\epsilon$}
\def\OH{{12\-+log(O/H)} }
\def\lNO{log(N/O) }
\def\cd{chemo-dynamical }
\def \rmaa {Rev.Mex.\allowbreak  Astron.\allowbreak  Astrofis.}

\title{Internal \cd Modeling of Gas Exchange within
Galaxies and with their Environment}   
\author{Gerhard Hensler$^1$, Simone Recchi$^{1,2}$, 
Danica Kroeger$^3$, Tim Freyer$^3$}   
\affil{
$^1$ Institute of Astronomy, University of Vienna,
T\"urkenschanzstr. 17, 1180 Vienna, Austria,
email: hensler@astro.univie.ac.at \\
$^2$ INAF - Osservatorio Astronomico di Trieste, 34131 Trieste, Italy \\
$^3$ Inst. of Theor. Physics {\rm \&} Astroph., U of Kiel, 
24098 Kiel, Germany
}

\begin{abstract} 

By a few but important examples as models of combined radiative and wind-driven 
\HII regions and galactic winds we demonstrate the 
importance of refined small-to-medium scale studies of \cd effects.
These processes determine the internal dynamics and energetic of the ISM 
and affect its observational signatures, e.g. by abundance contributions,
but are not yet reliably and satisfactorily explored. 
\end{abstract}

\vspace{-0.5cm}

\section{Introduction}   

Galaxies experience energy deposit and element release into their ISM
from massive stars during the lives and by their explosions.
They are main drivers of energetics and dynamics in galaxies and in
both stages they release characteristic elements which are preferably 
contained in the hot gas phase. Nevertheless, because of the
close coupling of hot, warm, and cool ISM energetic and dynamical processes
lead to \cd effects which alter the energy budget and the element
content in the different gas phases.
Understanding these processes from stellar to galactic scales are therefore
vital for understanding and modeling the evolution of galaxies and also
their interaction with the environment.

\section{Massive Stars}

\subsection{Effects of massive stars during their lives}

Massive stars play a crucial role in the evolution of galaxies, 
as they are the primary source of metals, and they 
dominate the turbulent energy input into the interstellar medium (ISM) by their massive 
and fast stellar winds, the ultraviolet radiation, and supernova explosions. 
The radiation field of these stars, at first, photo-dissociates the ambient 
molecular gas and forms a so-called photo-dissociation region (PDR) of neutral 
hydrogen. Subsequently, the Lyman continuum photons of the star ionize 
the \HI gas and produce a \HII region that expands into the neutral ambient medium. 

As these stars have short lifetimes of only a few million years, 
\HII regions indicate the sites of star formation (SF) and are targets to measure 
the current SF rate in a galaxy. Furthermore, the 
emission line spectrum produced by the ionized gas allows the accurate determination 
of the current chemical composition of the gas in a galaxy. 
Although the physical processes of the line excitation are quite well understood and 
accurate atomic data are available, so that the spectral analysis of \HII regions 
\citep[see e.g.][]{stas79,evdo87} serves as an essential tool to study 
the evolution of galaxies, their reliability as diagnostic tool have also to be 
studied with particular emphasis e.g.\ to temperature fluctuations
\citep{peim67,stas02} and line excitations.

While the simple concept of a uniform medium in ionization equilibrium with the
radiation from the stars (the Str\"omgren sphere) is successful in describing several
global features of the object and allows to predict the emission line spectrum quite
reliably, it has long been realized that \HII regions are complex and dynamical objects. 
Dynamical modelling of \HII regions caused purely by the energy deposit of the stellar 
radiation field has therefore been started already long ago 
\citep[see e.g.][and references therein]{york86} 
providing both a first insight into the formation of dynamical structures 
and allowing to derive more realistic scaling laws for observational comparisons, 
respectively.

That massive stars do not only release energy by their radiation but also act
by means of strong stellar winds which drive a turbulent ISM but, on the other hand, 
both are metal dependent has three major consequences for galaxies with present-day 
solar metallicities: 
at first, the classical \HII has lost its simplicity, because, 
secondly, it engulfs a vehemently expanding stellar wind bubble (SWB) and, 
thirdly, peels off the stellar outermost shells, by this, incorporating nuclear 
burning regions into the wind mass loss.

\subsection{Star formation Feedback}

Quiescent SF as a self-regulated process is a widely accepted concept.
The stellar feedback can adapt both signs, positive as a
triggering mechanism in a self-propagating manner like in superbubble
shells \citep{ehle97,fuha00} vs. negative as self-regulation.
Primarily the correlation between the surface density of disk galaxies' 
\HI gas and the vertically integrated SF rate derived from the \Ha flux
\citep{kenn98} serves as the best proof of a SF
self-regulation. Even more refined observations that trace more complex 
molecular gas \citep{gaso04} follow the same surface density 
relation. Surprisingly, this relation can be reproduced by different 
numerical models and with different prescriptions from purely collapsing 
cloud cores \citep{li05} via thermodynamical cloud models \citep{krke05} 
to sophisticated global \cd treatments of galaxy models \citep{burk92,harf06} 
and even seems to hold under the neglection of self-regulation processes 
in cosmological simulations \citep{krav03}. 
This fact obtrudes that the gas dependence of processes within the SF
matter cycle although expected to behave non-linearily cancels out and 
results in a universal law. Such a dependence is e.g.\ plausible because 
K\"oppen et al. (1995) 
have demonstrated that due to the square-dependence of collisionally
excited radiative cooling on the gas volume density also the SF
rate adapts to that over a large density range and almost independent of
the heating rate and other fiducial parameters. This also holds for
a multi-phase ISM accounting for supernova typeII (SNII) production 
of hot gas and for gas transitions by evaporation 
and condensation \citep{koep98}. From this result it seems reasonable that 
the vertical accumulation of gas mass should also lead to a SF rate
dependence on the column gas density because the self-gravitational potential 
determines the disk stratification. The Kennicutt law therefore could reflect 
a sum over the volumes, while the 1.4 power is also close to the pure
free-fall criterion.. 
Then the timescales must definitely tell us about both process and
carrier of self-regulation, respectively.

\subsection{Evolution and structure of bubble around massive stars}

However, this question needs more exploitation yet. Already the energy
deposit by massive stars described by a parameter \eps , called energy
transfer efficiency, is neither well derived during their normal lives
nor for their SNII explosions and the expanding SN remnants. 
The reason is that for both stages the environmental state and thus 
preceding processes as well as dynamical effects play a significant role.
Analytical estimates for purely radiative \HII regions yield \eps\ 
of the order of a few percent \citep{lask67}. 
Although the additional stellar wind power $L_w$ 
can be easily evaluated from model and observational values \citep{kupu00},  
its fraction that is transferred e.g.\ into 
thermal energy, i.e.\ the thermal \eps , or into turbulent energy is not 
obvious from first principles. The picture is that
the fast stellar wind creates shocks that form the SWB filled with very 
hot plasma, which expands into the \HII region so that also this \HII
shell has to expand into the surrounding ISM due to its overpressure.

The structure and evolution of SWBs and the transition to its surrounding 
\HII region can be described by a set of equations \citep{weav77} from which
the wind-energy deposit can be derived theoretically under the simplifying 
assumptions of a point source of a constant and spherically symmetric strong 
wind that interacts with a homogeneous ambient ISM. The \eps 's amount to 
significant fractions of $L_w$ \citep*[see][and references therein]{frey03}.
It has become evident that the stellar parameters such as 
the mass-loss rate, the terminal velocity, the effective temperature and the 
luminosity of the star vary strongly during the stellar lifetimes. While most 
previous studies dealt with either the evolution of \HII regions or of SWBs 
separately and with constant values, little is known about the interaction 
of these two structures.

Although the analytical and semi-analytical solutions for the evolution of
SWBs have been improved over the years as well as the numerical simulations 
have been done with increasing complexity, like e.g.\ 2D calculations of SWBs  
\citep[see e.g. by][and a series of papers]{rozy85}
and/or combined 1D radiation-hydrodynamical models of \HII region coupled 
with the dynamical SWB 
\citep[e.g. by][for references see Freyer et al. 2003]{gaml95},
a variety of physical effects remains to be included in order to
achieve a better agreement of models and observations with regard to the 
evolution of the hot phase in bubbles \citep{mclo00,chu00}.

\begin{figure}[!h]
\plottwo{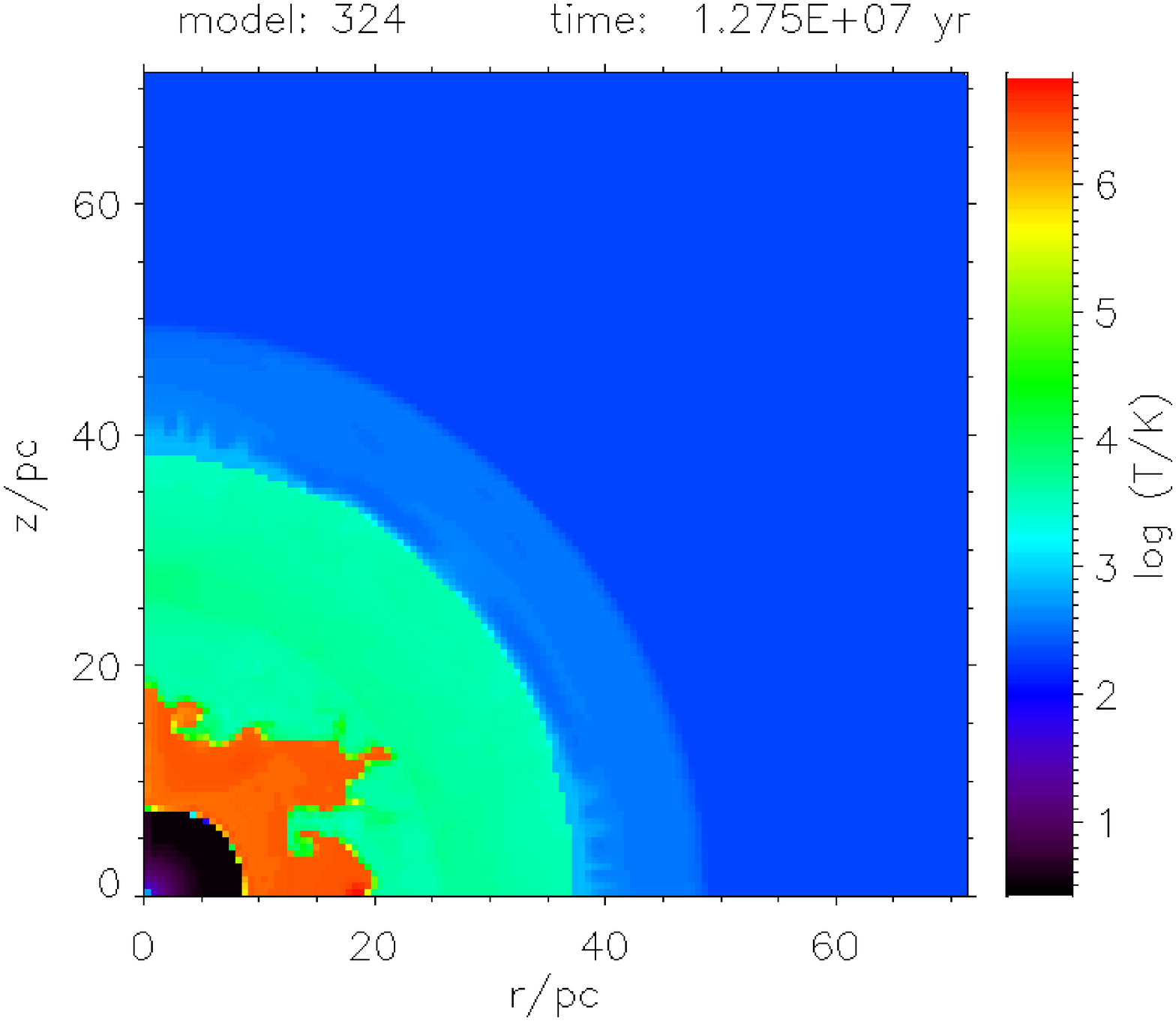}{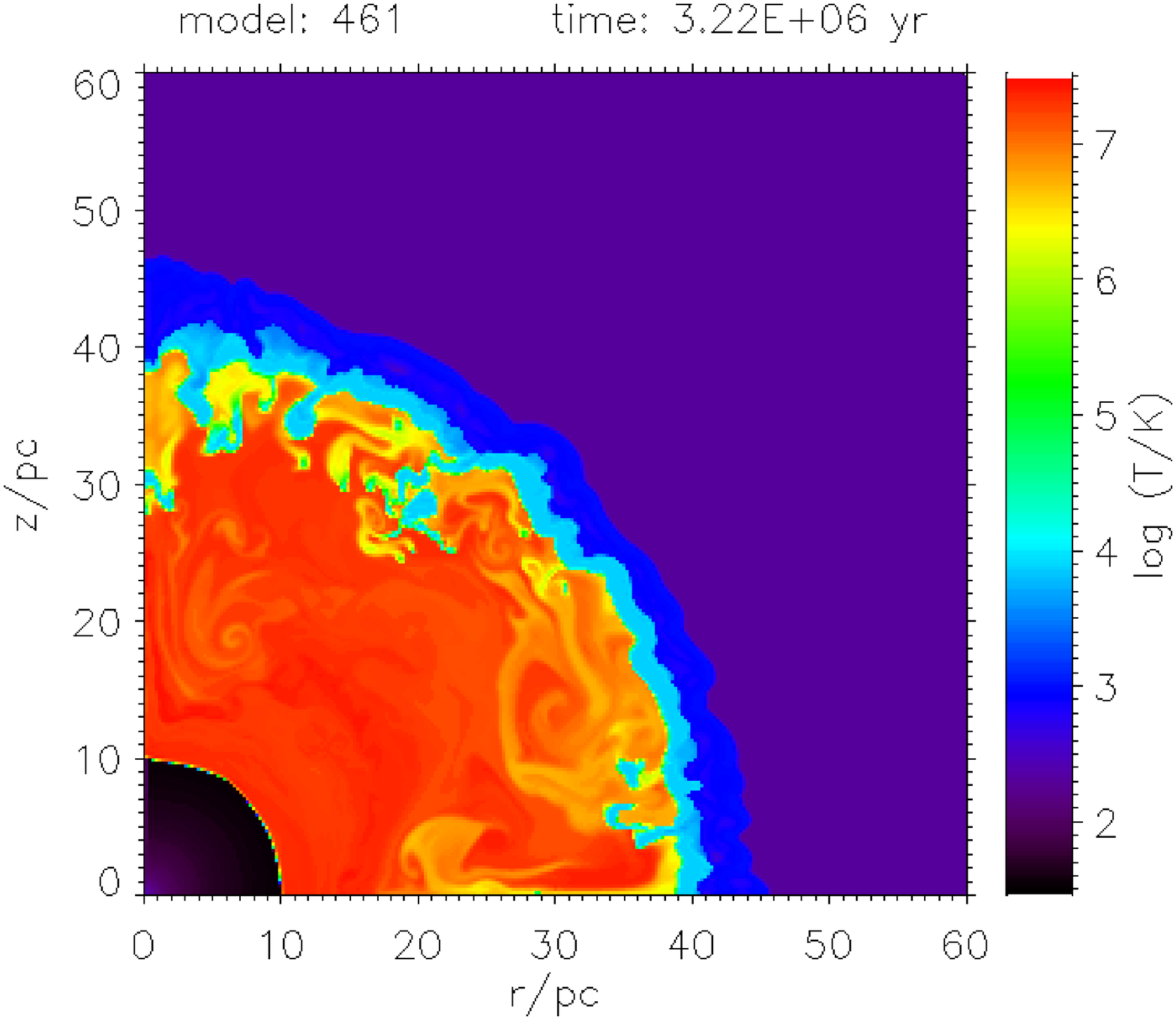}
\vspace{-0.3cm}
\caption{Temperature distributions of a 15 \Msun (left panel) and
a 85 \Msun (right panel) model at late stages of their stellar evolution.
The temperatures are colour coded to the right of the plots but differently
 (taken from Kroeger 2006). For discussion see text.}
\vspace{-0.5cm}
\end{figure}

To improve the insight into the evolution of radiation-driven + wind-blown
bubbles around massive stars, we have performed a series of
radiation-hydrodynamics simulations with a 2D cylindrical-symmetric
nested-grid scheme for stars of masses 15 \Msun (Kroeger et al. in prep.), 
35 \Msun \citep*{frey06}, 60 \Msun \citep*{frey03}, and 85 \Msun \citep{kroe07}. 
The main issues are:

\noindent 
1) The \HII region formed around the SWB has a complex structure
mainly affected by dynamical processes like e.g. shell instabilities,
vortices, mixing effects, etc. It is more compresssed by stronger winds 
(see Fig.1).

\noindent 
2) Finger-like and spiky structures of different densities and temperatures
are formed in the photo-ionized region \citep{frey03}.

\noindent 
3) The regions contributing to the \HII emission line are not solely
limited to the photo-ionized shell around the SWB but also form from 
photo-evaporated gas at the trailing surface of the SWB shock front 
(see 85 \Msun model in Fig.1). 

\noindent 
4) Because dispersion of this cooler photo-evaporated gas into the hot SWB 
leads to mixing also the stellar material expelled by the wind emerge partly 
in the spectra.

\noindent 
5) As a consequence the metal-enrichment of the wind in the Wolf-Rayet stage
which is generally assumed to remain only in the hot SWB for a long time affects
the observationally discernible abundance of the \HII gas.
By these models \citep{kroe06b} it could be proven for the first time that the 
metal release by Wolf-Rayet stars can be mixed within short timescales from the 
hot SWB into the warm ionized gas and should become observationally accessible.
As the extreme case for the 85 \Msun star we found a 22\% enhancement of Carbon,
but neglible amouts for N and O.

\noindent 
6) As expected from the distribution of \HII gas the radially projected
\Ha brightness shows a decrease to the center and a slight brightening to the 
limb but not as strong as expected according to the increase of the line-of-sight 
with impact parameter \citep{frey03}. This effect depends on the bubble
age and starts from central brightning. It also demonstrates both: 
the neglection of heat conduction and the homogeneous initial density
do not allow a sufficient brightning of heat conductive interfaces so that,
secondly, only the photo-evaporated backflow can contribute to the \Ha 
luminosity in present models. 
In reality, condensations which become embedded into the hot SWB are 
exposed to heat conduction.

\noindent 
7) The sweep-up of the slow red supergiant wind by the fast Wolf-Rayet 
wind produces remarkable morphological structures and emission signatures which
agree well with observed X-ray luminosity and temperature as well as with the
limb brightening of the radially projected X-ray intensity profile 
\citep[for details see][]{frey06}. 

\noindent 
8) \eps's for both radiative as well as kinetic energies remain much lower than 
analytically derived (more than one order of magnitude) and amount to only 
a few per mil \citep{hens07}. There is almost no dependence on the stellar
mass what is principly expected because the energy impact by Lyman continuum photons
and by wind luminosity increase with stellar mass. 
Vice versa, since the gas compression is stronger by a more energetic wind
also the energy loss by radiation is more efficient.

Nonetheless, a word of caution and unfortunately of discouragement has to be
expressed here because the stellar evolutionary models 
are not unique but depend on the authors. In order to get a quantifiable comparison
of the models by Garc\'{\i}a-Segura et al. (1996a, 1996b) 
with our 2D radiation-hydrodynamical
simulations (Freyer et al. 2003, 2006) for the 35 and 60 \Msun studies we used
the same stellar parameters. Since no stellar 
parameters were available from the same group for the 15 and 85 \Msun models
we had to make use of the Geneva models \citep{scha92}.
A comparison of the age-dependent 60 \Msun parameters of both groups has revealed
enormous differences in the energetics by almost one order of magnitude as
well as that the Wolf-Rayet and Luminous Blue Variable stages occur in 
contrary sequence, respectively \citep{kroe06a}. 

Furthermore, \HII regions show a great diversity of shapes which give direct evidence 
that the ISM is not a smooth, homogeneous, and uniform medium. 
In addition, the complexity of the internal structure and of the flow pattern of
the gas indicates that the evolution of the ionized region is strongly affected
by existing irregularities and hydrodynamical instabilities. 
And even more, stellar wind bubbles and SNRs of neighbouring stars collide
and may overlap to form superbubbles with diameters of several 100 pc scale, 
which are prominent by their X-ray emission in galaxies with active SF.
This picture of SWB/\HII aggregates has also to be extended by PDRs which are
mainly produced in the early evolutionary phases of ultra-compact \HII regions. 
Since they are heavily dust-enshrowded they emerge from GLIMPSE observations
\citep{benj03} and allow detailed diagnostics of these common stellar bubbles 
\citep{chur06}.

\section{Galactic Mass Exchange}

\subsection{Galactic Outflows}

Finally, massive stars explode as SNeII, creating a SN remnant that 
sweeps up the ambient medium. These phases have been understood 
and modeled in detail energeticly and dynamically. Cumulative SNeII form
superbubbles which expand vehemently and can drive a galactic outflow.
Since the hot superbubble gas carries SNII elements and has to act dynamically
with the surrounding ISM, it has huge energetic and chemial effects on galaxies.
Therefore, the questions that arise for their chemo-dynamical contribution to
galaxy evolution are the same as before for the \HII/SWB regions.
Although investigations have been performed for \eps\ of SNe \citep{thor98}
and superbubbles \citep[e.g.][]{stri04} they are yet too simplistic 
(e.g.\ only 1D) for quantitative results. 

Because of their low gravitational potential dwarf galaxies are sensitive to 
energetic processes like superbubbles so that these can easily develop to a 
galactic outflow as perceivible e.g.\ in NGC~1705 \citep{hens98}, 
NGC~1569 \citep{mart02}, I~Zw~18 \citep{mart96}, and many other Blue
Compact Dwarf Galaxies (BCDs).
Dynamical models \citep{mlfe99} can still provide only a qualitative understanding.
Nevertheless, the element abundances stemming from different stellar progenitors
should enable one to trace back the history of element enhancement vs. depletion 
by different processes. Since in general the SF varies in BCDs and, furthermore,
produces massive star clusters one
can imagine that a first superbubble has to exert most of its energy for its
expansion while the successive SNII explosions can be funneled through the 
chimney and by this more easily expel its metal-enriched gas. This can only happen,
if the next SF episode follows shortly enough after the former one. For this 
scenario it is, thus, important to explore the timescale of refill superbubbles vs.
the SF timescale. For the refill of superbubbles in the gas layer of typical
BCDs we \citep{rehe06} found a range of a few 100 Myrs. This means, that the
amount of galactic winds depends not only on the SF rate but also on its modes, 
bursting vs. gasping, but affects the \cd evolution \citep{hens02}.

\begin{figure}[!h]
\plotfiddle{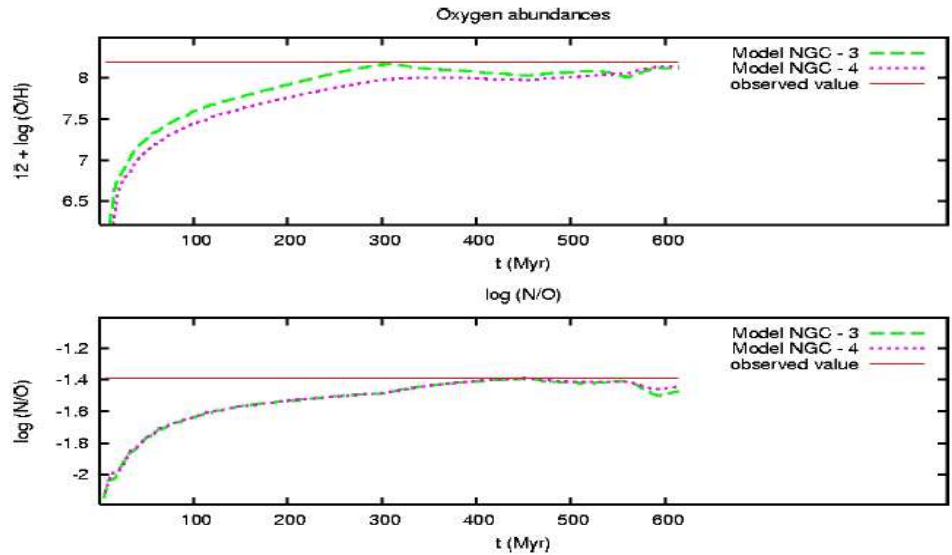}{7.5cm}{0}{55}{45}{-190}{10}
\vspace{-0.7cm}
\caption{Comparison of abundance evolution in 2 modls representing
NGC~1569: NGC-3 with a single strong and short SF burst, 
NGC-4 with 2 former SF epochs and a mild recent burst. 
\citep[For details see][]{rehe06}.}
\vspace{-0.2cm}
\end{figure}

For the best studied BCDs, I~Zw~18 \citep{recc04} and NGC~1569 \citep{recc06},
2D dynamical models with characteristic tracer elements have been performed 
aiming at achieving a best fit with observations. 
The main results can be briefly summarized as follows:
Galactic winds always occur.
Models with bursting SF are generally unable
to account for the chemical and morphological properties of NGC~1569,
because they either underproduce O or inject too much energy into the 
system, enough to loose a too large gas fraction. The best agreement
with the chemical composition is found for long-lasting continuous episodes
of SF of some hundred Myrs of age and a recent more intensive short SF burst
(see Fig.2). In most models with gasping SF, the final chemical composition 
of the galaxy reflects mostly the chemical enrichment from old stellar 
populations. In fact, if the first episodes of SF are powerful enough to create
a galactic wind or to heat up a large fraction of the gas surrounding the 
SF region, the metals produced by the last burst of SF are 
released into a too hot medium or are directly ejected from the galaxy 
through the wind. They do not have the chance to pollute the surrounding 
medium and to contribute to the chemical enrichment of the galaxy.
This result confirms the conclusion that most of the O from the
last SF epoch is stored in the hot X-ray gas \citep{mart02}.

\subsection{Gas Infall}

Although the element abundance properties of most BCDs favour only a 
young stellar population of at most 1-2 Gyrs, they mostly consist of 
an underlying old population as it is the case for NGC~1569.
Furthermore, most of the objects are embedded into \HI envelops from
which at least NGC~1569 definitely suffers gas infall 
\citep{stis02,mueh05}. This has lead K\"oppen \& Hensler (2005) 
to exploit the influence of gas infall with metal-poor gas into an 
old galaxy with continuous SF on particular abundance patterns.
Their models could match not only the observational regime of BCDs in 
the [\OH-\lNO] space but also explain the shark-fin shape of their data
distribution by means of the ratio of mass infall to existing cloud mass
in the sense that for larger infall fractions the loop is more extended.
 
Analytical investigations as well as too simplified numerical models, however,
are unable to distinguish between the different gas phases and their metal
content in order to derive abundances in the targets of observations.

In addition, also mixing effects at interfaces between gas phases and due to
turbulence (like e.g. those in the combined SWB/\HII complexes)
contribute to the observations by the enhancement or, respectively, dilution 
of metals. Since gas infall is not only affecting the chemistry but also SF 
\citep{hens04} and dynamics of outflows we have extended 
the former models (Recchi et al. 2004, 2006) by infalling clouds \citep{rehe07}. 
In these models, two kinds of cloud contributions are considered, only initially
existing and continuously formed, respectively. The issues are the following: 
Due to dynamical processes and thermal evaporation, the clouds survive only 
a few tens of Myr. The internal energy of cloudy models is typically reduced 
by 20 -- 40 \% compared to models with a smooth density distribution.  
The clouds delay the development of large-scale outflows, helping
therefore in retaining a larger amount of gas inside the galaxy.  
However, especially in models with continuous creation of infalling clouds, 
their ram pressure pierce the expanding supershells so that through these holes
freshly produced metals can more easily escape and vent into
the galactic wind. Moreover, assuming for the clouds a pristine chemical 
composition, their interaction with the superbubble dilute the hot gas, reducing 
its metal content.  The resulting final metallicity is therefore, in general,
smaller (by $\sim$ 0.2 -- 0.4 dex) than the one attained by smooth models.

\acknowledgements 
The authors gratefully acknowledge contributions and discussions by
Jay S. Gallagher, Stefan Hirche, Joachim Koeppen, Andreas Rieschick, 
Christian Theis, Wolfgang Vieser, and Harald W. Yorke. 
Part of the work was funded by the DFG under grants HE~1487/17, /25, /28. 
G.H. cordially thanks Johan Knapen and the organizers for the invitation to 
this conference. The participation was made possible by grants from the 
conference and from the University of Vienna under BE518001.


\end{document}